\documentclass[amsmath,amssymb,11pt]{article}
\usepackage{jheppub2}
\pdfoutput=1
\usepackage{graphicx,epsfig}
\usepackage{float}
\usepackage{subfig}
\usepackage{subfloat}
\usepackage[utf8]{inputenc}
\usepackage{hyperref}
\usepackage{caption}
\usepackage[dvipsnames]{xcolor}
\usepackage{cleveref}

\usepackage{lmodern}

\newcommand{\beq}{\begin{equation}}

\newcommand{\eeq}{\end{equation}}

 \newcommand{\be}{\begin{equation}}
 \newcommand{\ee}{\end{equation}}
 \newcommand{\bea}{\begin{eqnarray}}
 \newcommand{\eea}{\end{eqnarray}}

\usepackage{tikz}
\usetikzlibrary{positioning}
\usetikzlibrary{intersections}
\usetikzlibrary{fadings} 
\usetikzlibrary{arrows.meta} 
\usetikzlibrary{arrows}

\tikzfading[name=fade out,
inner color=transparent!0,
outer color=transparent!100]

\definecolor{cherryblossompink}{rgb}{1.0, 0.72, 0.77}
\definecolor{lightblue}{rgb}{0.68, 0.85, 0.9}

\usetikzlibrary{decorations.pathmorphing}
\usetikzlibrary{decorations.pathreplacing,decorations.markings}

\usetikzlibrary{backgrounds,automata}

\title{Classical-quantum oscillators as diffusive processes in phase space}

\author{Emanuele Panella}
\emailAdd{emanuele.panella.21@ucl.ac.uk}

\affiliation{
Department of Physics and Astronomy, University College London,\\
Gower Street, London, WC1E 6BT, United Kingdom}

\abstract{The dynamics of hybrid systems -- i.e. ones in which classical and quantum degrees of freedom co-exist and interact -- feature both diffusion in the classical sector and decoherence in the quantum state. In this article, we will consider the simple setup of a classical damped oscillator interacting with its quantum counterpart and show that, for any initial state, the dynamics flows to a unique (non-equilibrium) steady state, which we compute explicitly. To do so, we make use of a useful mapping between hybrid and classical diffusive dynamics, which we characterise in detail in the master equation formalism.}

\begin{document}

\maketitle

\section{Introduction} \label{sec:intro}
Recent years have seen a resurgence in the interest for hybrid systems -- i.e. physical settings in which both classical and quantum degrees of freedom coexist and, crucially, interact. The main lesson from these investigations is that consistent hybrid dynamics is possible, but it requires both stochasticity in the classical degrees of freedom as well as loss of coherence in the quantum state~\cite{oppenheim2023gravitationally}.

Despite the original motivation is rooted in fundamental high-energy physics and gravitational physics~\cite{oppenheim2021postquantum}, classical-quantum (CQ) models are much more flexible and can be adapted as effective theories in order to describe the classicalisation of a fundamentally quantum subsystem~\cite{layton2024classical}. Natural settings in which the border between quantumness and classicality blurs -- other than the aformentioned semiclassical gravity and cosmology ~\cite{albrecht1994inflation,braden2019new-semiclassical} -- are quantum chemistry~\cite{tully1990molecular,c.-tully1998mixed}, continuous measurement-and-feedback~\cite{lloyd2000coherent,annby-andersson2022quantum} and condensed matter systems~\cite{steel1998dynamical,mink2023collective}.

The natural first step in exploring this novel framework would of course be to solve for the simplest system at hand, a classical oscillator interacting with a quantum one. The first treatment of the hybrid oscillator appeared in~\cite{sajjad2025a-quantum}, by use of the path-integral description of CQ. There, the dynamics considered did not include any mechanism to remove energy from the system. As a consequence, due to the combination of diffusion and decoherence -- which on average inject energy -- the two coupled oscillators heat up indefinitely. A more recent work~\cite{layton2025restoring} studied the most general dynamics for the hybrid systems that preserves the CQ thermal state. In the case of a CQ oscillator, it was found that the evolution needs, other than friction in the classical sector, a specific (temperature dependent) type of Lindblad operators controlling the dynamics of the quantum subsystem~\cite{layton2025restoring}. 

In this article, we revisit the coupled classical-quantum system of oscillators, starting from the same dynamics of~\cite{sajjad2025a-quantum} with regards to the reversible terms -- but including friction in the classical sector. The damping in the classical oscillator is natural if one sees this model as an effective one. For example, consider two quantum oscillators, one of which is interacting with a thermal bath~\cite{syu2021entanglement}. The interaction with the environment can induce classicalisation in that half of the bipartite system~\cite{layton2024classical}, and would be origin of both the diffusion and dissipation. As we will see, our hybrid oscillator reaches a steady state after some characteristic timescale associated with the damping coefficient, although the system does \emph{not} approach equilibrium, unlike in~\cite{layton2025restoring}. By this, we mean that the fluctuation-dissipation relations are not satisfied -- this implies that although the property of the end-state of the evolution are time-dependent, there is some non-zero entropy production in the system~\cite{seifert2012stochastic}. This setup therefore offers a useful toy model under analytical control to study non-equilibrium thermodynamics in CQ.

The plan is as follows. In Section~\ref{sec:theory} we review the path integral and master equation representation of CQ dynamics. The main results are in Section~\ref{sec:results}. We first analyse in Section~\ref{sec:classical} the simpler case of two stochastic classical coupled oscillators, of which only one is damped. We recognise the equations of motion as a multidimensional Ornstein-Uhlenbeck (OU) process and, using standard results, we prove that the coupled oscillators univocally approach a steady state -- of which we compute equal-time covariances. We then move to the path integral representation of the process, which we use to find the unequal-time covariances between the position of the oscillators (we report the explicit results for the small-coupling regime). In Section~\ref{sec:quantum} we quantise the non-dissipative oscillator. We use the similarity between the classical Martin-Siggia-Rose path integral and the quantum Schwinger-Keldysh functional in the average-difference basis to map the classical result to the hybrid oscillator -- from which we extract its statistical two-point functions. We conclude by analysing the relation between the classical and CQ dynamics more in detail in Section~\ref{sec:phase_space}. There, we perform the Wigner-Moyal transform to the CQ master equation to map the evolution of the hybrid state in phase-space. There, we show explicitly that, for interaction potential that are at most harmonic, CQ dynamics can be exactly represented as a classical diffusion process.

Note that throughout the article we use the definition of thermal state as being the canonical Gibbs' state with respect to the Hamiltonian of the system $H_{CQ}$. That is, the same Hamiltonian generates the equal-time and unequal-time correlations for the system. This is in contrast with a looser definition of thermal state, which we do not adopt here, where the Hamiltonian appearing in the expression for the Gibbs state is allowed to be not the system's, but some other -- e.g. the “mean-field hamiltonian” --which usually has to depend on the temperature of the state itself~\cite{trushechkin2022open}.

\section{Background} \label{sec:theory}
We now quickly brush over two equivalent representations of hybrid classical-quantum dynamics -- the master equation formalism~\cite{oppenheim2021postquantum,oppenheim2022classes} and the path integral approach~\cite{oppenheim2021postquantum,oppenheim2023path,oppenheim2023covariant}. For a comprehensive review, please refer to~\cite{weller-davies2024classical-quantum,diosi2024classical-quantum}, as here we will focus on a restricted family of allowed evolution -- those that are “Hamiltonian”, in a sense that we clarify shortly, and those that saturate the decoherence-diffusion trade-off. The latter is a consistency condition, key in hybrid dynamics, that forces the evolution to be non-unitary in the quantum sector and stochastic in the classical subsystem. By saturating the trade-off, we mean that we consider dynamics with minimal decoherence for a given classical diffusion tensor.

\subsection{CQ master equation}
We will be interested in dynamics that can be seen as generated by some interaction Hamiltonian 
\begin{equation}
    H_{CQ} = H_C (q,p) +H_Q(Q,P)+ V_I(q,Q) 
\end{equation}
coupling the classical and the quantum system, i.e. the minimal form of consistent CQ dynamics that, modulo the irreversible terms, corresponds to the standard mean-field semiclassical evolution. Specifically, $H_C$ and $H_Q$ are the Hamiltonians of the classical and quantum systems respectively, $(q,p)$ and $(Q,P)$ their phase-space variables, and $V_I$ an interaction potential coupling only to the generalised positions of the two systems. The evolution of the CQ state $\varrho$, a subnormalised quantum density matrix with classical phase-space dependence obeying 
\begin{equation}
    \int_{\mathcal{M}} dq \ dp \ \mathrm{Tr}_\mathcal{H}[\varrho] = 1 \ ,
\end{equation}
where $\mathcal{M}$ is the classical phase-space and $\mathcal{H}$ the Hilbert space of the quantum subsystem, is highly constrained by the requirements of complete positivity, linearity, trace-preservation and Markovianity. Whilst the latter is a modelling assumption (and a great simplification), the other constraints are imposed to retain the statistical interpretation of the hybrid state.

Then, by minimally coupling the classical system to the noise field (i.e. modelling it as a stochastic white noise force) and choosing as basis for the Lindblad operators $L_i = \frac{\partial V_I}{\partial q^i}$, the general master equation greatly simplifies too~\cite{oppenheim2021postquantum,oppenheim2022classes}
\begin{equation}
\begin{split}
\label{eq:CQ_master}
    \frac{\partial \varrho(q,p)}{\partial t} &= \{H_C, \varrho \} + \frac{1}{2}\frac{\partial^2}{\partial p_i \partial p_j} \left(D_{2,ij}(q) \varrho \right) +\frac{1}{2}\left( \{V_I,\varrho\}-\{\varrho,V_I^\dagger \}\right) \\
    & \qquad- i [H_Q+V_I, \varrho] + D_0^{ij} \left(\frac{\partial V_I}{\partial q_i} \varrho \frac{\partial V_I^\dagger}{\partial q_j}- \frac{1}{2}\left\{ \frac{\partial V_I^\dagger}{\partial q_j}\frac{\partial V_I}{\partial q_i}, \varrho\right\}_+\right)  
    \\
    &= \{H_{CQ}, \varrho \}_A + \mathbf{D}[\varrho] \ ,
\end{split}
\end{equation}
where we have grouped the reversible part of the dynamics in the Aleksandrov bracket
\begin{equation}
    \{H_{CQ}, \varrho \}_A \equiv \{H_C, \varrho \} - i [H_Q+V_I, \varrho] +\frac{1}{2}\left( \{V_I,\varrho\}-\{\varrho,V_I\}\right)
\end{equation}
and the irreversible part in the decoherence-diffusion operator
\begin{equation}
    \mathbf{D}[\varrho] \equiv \frac{1}{2}\frac{\partial^2}{\partial p_i \partial p_j} \left(D_{2,ij}(q) \varrho \right)  + D_0^{ij} \left(\frac{\partial V_I}{\partial q_i} \varrho \frac{\partial V_I^\dagger}{\partial q_j}- \frac{1}{2}\left\{ \frac{\partial V_I^\dagger}{\partial q_j}\frac{\partial V_I}{\partial q_i}, \varrho\right\}_+\right) \ .
\end{equation}

\subsection{CQ path integral}
The time-local dynamics can be trotterized and, therefore, expressed in terms of integration over paths. The deterministic part of the dynamics of hybrid system can be derived from the CQ \emph{proto-action}: 
\begin{equation}
    W_{CQ} = \int \mathrm{d}t \left(L_Q[Q]+L_C[q]-V_I[q,Q] \right) = \int\mathrm{d}t \ L_{CQ}\ .
\end{equation}
The prefix “proto” here indicates that, whilst the functional encodes all the information about the deterministic part of the dynamics, it is not the action of the path-integral itself. Instead, continuous hybrid dynamics of the form of Equation~\ref{eq:CQ_master} involving $q$ and $Q$ (the classical and quantum degree of freedom respectively) can be equivalently represented via the following configuration-space path integral~\cite{oppenheim2023path,oppenheim2023covariant} (assuming the Lagrangians are quadratic in the momenta, such that they can be trivially integrated out):
\begin{equation}
    \varrho(q,Q^L,Q^R,T)= \int \mathcal{D}Q^{L/R}\int\mathcal{D}q \int\mathcal{D}\tilde{q} \ e^{iI_{CQ}} \varrho(q,Q^L_0,Q^R_0,t_0) \ ,
\end{equation}
where
\begin{equation}
\begin{split}
\label{eq:CQ_action}
    iI_{CQ} = \int \mathrm{d}t &\left[i\Delta L_{CQ}[q,Q^{L/R}]- \frac{1}{2} \frac{\delta\Delta W_{CQ}}{\delta q^i} D_0^{ij}\frac{\delta\Delta W_{CQ}}{\delta q^j} \right. \\
   & \qquad \left.- \tilde{q}^i \frac{\delta}{\delta q^i}\bar{W}_{CQ}[q,Q^{L/R}] + \frac{1}{2}\tilde{q}_i (D^{-1}_2)^{ij}\tilde{q}_j \right] \ .
\end{split}
\end{equation}
For simplicity of notation we have defined the averaged and subtracted proto-actions:
\begin{equation}
    \Delta W_{CQ} = W_{CQ}[q,Q^L]-W_{CQ}[q,Q^R] \ ,\qquad \bar{W}_{CQ} = \frac{1}{2}\left( W_{CQ}[q,Q^L]+W_{CQ}[q,Q^R] \right) \ .
\end{equation}

The hybrid path integral neatly splits up in a quantum Fenynman-Vernon (FV) and a Martin-Siggia-Rose (MSR) term, the first and second lines in Equation~\ref{eq:CQ_action} respectively. The FV path integral is a generalisation of the unitary quantum path integral, in which the quantum degrees of freedom are doubled into a left and right branch (indicated by a $L$ and $R$ respectively) in order to allow for unitary-breaking terms (encoded by left-right interaction terms in the action). On the other hand, the MSR path integral is an equivalent representation of stochastic dynamics, in which an auxiliary, imaginary, “response variable” (indicated by a tilde) is introduced to model the stochastic kicks. Useful references for the interested readers are~\cite{sieberer2016keldysh,hertz2016path}. The usual path integral techniques to compute expectation values of operators apply.

\section{Main results} \label{sec:results}
\subsection{The classical case}
\label{sec:classical}
We begin by considering two coupled -- stochastically driven -- classical oscillators. As we argue in Section~\ref{sec:quantum} and show explicitly in Section~\ref{sec:phase_space}, the classical-classical result can be mapped exactly to the classical-quantum result with the proper dictionary -- plus it has the virtue of being straightforward to interpret. 

The deterministic, conservative, part of the dynamics is encoded by the classical Hamiltonian
\begin{equation}
    H = H_1 + H_2 +V_I \ ,
\end{equation}
where $H_i = p_i^2/2m_i + \kappa_i q_i^2/2$ are the free Hamiltonian of the $i$-th oscillator, whilst the interaction potential is given by $V_I = \lambda (q_1-q_2)^2/2$. Here $m_{i}$ and $\kappa_{i}$ are the mass and the spring constant of the $i$-th oscillator whilst $\lambda$ is the coupling constant between the two particles. We damp the first oscillator with friction, with damping coefficient $\alpha$. The resulting (stochastic) equations of motion are given by
\begin{equation}
\label{eq:class_eom}
\begin{split}
    m_1\ddot{q}_1 + \kappa_1 q_1 + \alpha\dot{q}_1 + \lambda(q_1-q_2) &= \sqrt{D_1} \xi_1 \\
    m_2\ddot{q}_2 + \kappa_2 q_2 + \lambda(q_2-q_1) &=\sqrt{D_2} \xi_2 \ ,
\end{split}
\end{equation}
The stochastic forces $\xi_{1,2}$ are white noise processes obeying:
\begin{equation}
    \mathbb{E}[\xi_i(t)] = 0  \ , \qquad \mathbb{E}[\xi_i(t) \xi_j(t')] = \delta_{ij} \delta(t-t') \ ,
\end{equation}
meaning they are two mean-zero independent processes of unit variance. Physically, we are driving the $i$-th mass with independent random kicks of typical size $\sqrt{D_i}$. 

Note that these equations of motion correspond to two classical interacting oscillators in contact with separate heat baths -- one of which (the one exchanging energy with the frictionless mass) is an infinite-temperature bath. Indeed, the Einstein relation tells us that for a finite diffusion coefficient $D$, the friction coefficient $\alpha$ on the system of interest vanishes as the temperature of the bath diverges~\cite{marconi2008fluctuationdissipation:}. Chains of quantum or classical oscillators in contact with reservoirs at different (although both finite) temperatures have been already studied in the literature, and they naturally lead to non-equilibrium steady-states~\cite{rieder1967properties,zurcher1990quantum-mechanical,syu2021entanglement}. Although our system is quantitatively different, we expect (and we show it to be indeed the case) that it also flows to a non-equilibrium steady-state for arbitrary initial conditions.

Formally, this is an Ornstein-Uhlenbeck (OU) process -- a multidimensional stochastic process of the form:
\begin{equation}
\label{eq:OU}
    \dot{z}^i = - \Theta^i_j z^j  + \Sigma^i_j  \ \xi^j_t \ ,
\end{equation}
where $z^i$ are the components of the vector representing the degrees of freedom of the system whilst $\mathrm{d}W_t^i$ is a vector of independent white noise processes -- here we have assumed, without loss of generality, that the two have the same dimensions. The constant matrices $\Theta^i_j$ and $\Sigma^i_j$ encode the mean dirft of $z^i$ and covariance of the stochastic kicks respectively. An important result, which we will refer to over and over, is that for a multidimensional OU process, a steady-state exists if and only if the deterministic system is strongly stable -- i.e. the eigenvalues $\theta_i$ of $\Theta^i_j$ have strictly positive real parts~\cite{godreche2018characterising}. 

The reader versed in the theory of stochastic processes will know that Equation~\ref{eq:OU} is formally undefined without specifying a choice for stochastic calculus. In this article, we intend all stochastic integration in It\^o sense -- although the fact that the covariance matrix in our equations of motion is $z$-independent makes all the choices of discretisation equivalent. Please, refer to Appendix~\ref{app:uncoupled_osc} for the treatment of the two uncoupled (damped and non-dissipative) oscillators -- it is a good warm-up, with straightforward application of the main results from the theory of stochastic processes that we use in the next Section to prove the existence of the steady state -- including a mention of It\^o's lemma. 

\subsubsection{The steady-state}
We now show that the two oscillators reach a steady-state for non-zero coupling $\lambda$. Evidence that this is the case is obtained by looking at the average evolution of the energy under Equation~\ref{eq:class_eom}. Using the chain rule for stochastic processes, i.e. It\^o's lemma (see Appendix~\ref{app:uncoupled_osc} for a brief review), we straightforwardly obtain:
\begin{equation}
\label{eq:energy_growth}
    \mathbb{E}[\dot{H}] = -\frac{\alpha}{m_1^2}\text{Var}(p_1)+\frac{D_1}{2m_1}+\frac{D_2}{2m_2} \ ,
\end{equation}
meaning that the energy stops increasing once the variance in the momentum of the damped oscillator reaches
\begin{equation}
    \text{Var}(p_1) = \frac{1}{2\gamma_1}\left(D_1 + \frac{m_1}{m_2} D_2 \right) \ .
\end{equation}

To make sure the steady-state exist, however, we can refer to the aforementioned standard result from the theory of stochastic processes, and compute the stability of the system. First, note that
\begin{equation}
\label{eq:mat_coupl}
    \mathbf{\Theta} = \begin{pmatrix}
0 & -\frac{1}{m_1} & 0 & 0 \\
\kappa_1+ \lambda & \frac{\alpha}{m_1} & -\lambda  & 0 \\
0& 0& 0 & -\frac{1}{m_2} \\
-\lambda & 0 & \kappa_2+ \lambda  & 0 
\end{pmatrix} 
\ , \qquad
    \mathbf{\Sigma} = \begin{pmatrix}
0 & 0 & 0 & 0 \\
0 & \sqrt{D_1} & 0 & 0  \\
0 & 0 & 0 & 0 \\
0 & 0 & 0 & \sqrt{D_2}
\end{pmatrix} \ .
\end{equation}
We need to show that the eigenvalues of $\mathbf{\Theta}$ are all all strictly positive. To prove it, consider the eigenvalue equation:
\begin{equation}
\label{eq:eig_eq}
    \theta^4 - \gamma_1 \theta^3 + \left(\omega_1^2 +\omega_2^2 + \frac{\lambda}{m_1}+\frac{\lambda}{m_2} \right) \theta^2- \gamma_1 \left( \omega_2^2+\frac{\lambda}{m_2}\right)  \theta+\omega_1^2 \omega_2^2 + \omega_2^2\frac{\lambda}{m_1}+ \omega_1^2\frac{\lambda}{m_2} = 0 \ .
\end{equation}

The roots of the characteristic polynomial $P(\theta)$ are analytically solvable for being the solutions of a quartic equation. However, these solutions are extremely complicated expressions in general, so extracting them and requiring positivity of their real part is a very inefficient strategy. Instead, we make use of the Routh-Horowitz criteria~\cite{routh1877a-treatise,hurwitz1895ueber}, i.e. a series of criteria that need to be satisfied for all the solutions of a polynomial of order $n$ to have positive (or, equivalently, negative) real parts. These are more intuitive for the latter case, so we consider $\theta \to -\theta$ and prove that $P(-\theta)$ has solutions with only real negative parts. The first four conditions are equivalent to requiring that all the coefficients of the quartic are positive, which is always true if $\lambda,\gamma_1 >0$ -- a trivial condition. The remaining two criteria can be easily shown to reduce to:
\begin{itemize}
    \item $\left(\omega_2^2+\frac{\lambda}{m_2} \right)^2+\frac{\lambda^2}{m_2^2}>0$\ ,
    \item $\frac{\lambda^2}{m_2^2}>0$ \ ,
\end{itemize}
both trivially satisfied for real couplings. This shows that all the eigenvalues $\theta_i$ have positive real parts, meaning that the system reaches always a steady-state.

The reason for the existence of the steady-state for any coupling $\lambda$ is clearer when we explicitly solve (in perturbation theory) for the eigenvalues of the system -- we will see that the undamped oscillator develops an effective damping coefficient of order $\lambda^2$ due to the interaction. Physically, however, one can see that this has to be the case by a simple thermodynamics argument. The rate at which the energy is added into the system (for both oscillators) is fixed, and depends solely on the diffusion coefficient. However, the energy is extracted by the damped oscillator at a rate that depends on its typical velocity (and, hence, amplitude of oscillation). Since there is energy exchange between the two oscillators, the damped one will heat up until it reaches the typical size of the swings for which it ejects energy at the rate equal to the one at which it is being added to the combined system.

For an OU process, if the steady-state exists then it is Gaussian~\cite{godreche2018characterising}:
\begin{equation}
    P_{st}=(2\pi)^{-N/2} \mathrm{det}(\mathbf{C}_{\infty})^{-1/2} \exp\left(\frac{1}2{}z_i (C_{\infty}^{-1}) ^i_j z^j\right)
\end{equation}
The equal-time covariance of the OU process $\mathbf{C}_{\infty \ j}^i=\text{cov}(z^i,z^j)$ in such a state can be computed from the Lyapunov equation~\cite{godreche2018characterising}:
\begin{equation}
\label{eq:lyapunov}
    \mathbf{\Theta}\mathbf{C}_{\infty} + \mathbf{C}_{\infty} \mathbf{\Theta}^T = \mathbf{\Sigma} \mathbf{\Sigma}^T \ .
\end{equation}

Solving the Lyapunov equation, we find that the variance of the positions and momenta of the two oscillators are given by:
\begin{align}
    \mathbb{E}[p_1^2] &= \frac{1}{2\gamma_1}\left(D_1 + \frac{m_1}{m_2} D_2 \right) \label{eq:lyapunov_solp1p1} \\
        \mathbb{E}[p_2^2] &=  \frac{1}{2\gamma_1} \left[D_2\left(1 + \frac{m_1 m_2}{\lambda^2} \left(\left(\omega_1^2-\omega_2^2 + \frac{\lambda}{m_1} - \frac{\lambda}{m_2} \right)^2 + \gamma_1^2 \left(\omega_2^2 +\frac{\lambda}{m_2} \right)\right) \right)\right.\\
        & \hspace{50pt
        }\left.+\frac{m_2}{m_1} D_1 \right]  \notag \\
    \mathbb{E}[q_1^2] &= \frac{1}{2\gamma_1} \frac{D_1\left(\frac{\lambda}{m_2}+\omega_2^2\right)+\frac{m_1}{m_2}D_2\left(\frac{\lambda}{m_1}+\omega_1^2\right)}{m_1^2 \left(\omega_2^2 \frac{\lambda}{m_1}+\omega_1^2 \left(\omega_2^2 + \frac{\lambda}{m_2} \right) \right)} \\
    \mathbb{E}[q_2^2] &= \frac{1}{2\gamma_1} \frac{1}{m_2^2 \left(\omega_2^2 \frac{\lambda}{m_1}+\omega_1^2 \left(\omega_2^2 + \frac{\lambda}{m_2} \right) \right)}\left[\frac{m_2}{m_1}D_1 \left(\omega_1^2+\frac{\lambda}{m_1}\right) \right. \label{eq:lyapunov_solx2x2} \\
    & \hspace{50pt}\left.+D_2\frac{m_1m_2}{\lambda^2}\left(\left(\omega_1^2 +\frac{\lambda}{m_1}\right)^3+ \left(\omega_1^2 \left( \omega_2^2+\frac{\lambda}{m_2}\right) +\omega_2^2 \frac{\lambda}{m_1}\right) \right. \right. \notag \\
    & \left. \left. \hspace{50pt} \times \left(\omega_2^2 -2\omega_1^2+\frac{\lambda}{m_2}-2\frac{\lambda}{m_1}+\gamma_1^2\right)\right) \right] \notag  \ ,
\end{align}
Whilst the non-zero covariances are given by:
\begin{align}
    \mathbb{E}[p_1p_2] &= \frac{D_2}{2\gamma_1} \frac{m_1}{\lambda} \left( \omega_1^2-\omega_2^2+\frac{\lambda}{m_1}-\frac{\lambda}{m_2}\right)
     \label{eq:lyapunov_solp1p2} \\
    \mathbb{E}[q_1p_2] &= -\frac{D_2}{2\lambda} \\
    \mathbb{E}[q_2p_1] &= \frac{D_2}{2\lambda} \frac{m_1}{m_2} \\
     \mathbb{E}[q_1 q_2] &= \frac{1}{2\gamma_1}\frac{1}{m_2 m_1} \frac{D_1 \frac{\lambda}{m_1}+D_2 \frac{m_1}{\lambda}\left(\left(\omega_1^2+\frac{\lambda}{m_1}\right)^2 -\omega_2^2 \frac{\lambda}{m_1}-\omega_1^2 \left(\omega_2^2+\frac{\lambda}{m_2} \right)\right) }{\omega_2^2\frac{\lambda}{m_1}+\omega_1^2\left(\omega_2^2+\frac{\lambda}{m_2} \right)} \label{eq:lyapunov_solx1x2} 
\end{align}

The steady state variances have been checked numerically for a range of parameters -- the stochastic differential equations describing the trajectories of the system in phase space can be straightforwardly simulated with an Euler-Maruyama forward scheme~\cite{delia2024stochastic}.
    
\subsubsection{MSR path integral}
We will now take another, more generalisable route, to extract the unequal time two-point functions, by studying the MSR path integral of the process. As we will see, this is exactly solvable in theory, but requiring the roots of a quartic with general coefficients the exact solution is not illuminating. We will therefore work in the small $\lambda$ regime for the rest of the chapter and look for a more informative -- although approximate -- result.

The MSR path intregral representation of the stochastic process in Equation~\ref{eq:class_eom} (integrating out the momenta and working in configuration space) is given by:
\begin{equation}
    P(\underline{q}_T) = \int\mathcal{D}\underline{q} \int\mathcal{D}\underline{\tilde{q}}e^{-S[\underline{q},\underline{\tilde{q}}]}P(\underline{x_0}),
\end{equation}
where $\underline{q}=(q_1,q_2)^T$ is the vector state of the system with the position of the oscillators, whilst $\underline{\tilde{q}}=(\tilde{q}_1,\tilde{q}_2)^T$ are the so-called (purely imaginary) response variables. The MSR action for the process is given by:
\begin{equation}
\begin{split}
        S[\underline{q},\underline{\tilde{q}}] = \int_{t_0}^T dt& \left[ \tilde{q}_1\left(m_1 \partial^2_t+\kappa_1+\alpha\partial_t+\lambda \right) q_1 - \frac{D_1}{2}\tilde{q}_1^2 -\lambda\tilde{q}_1q_2 +\right.  \\
        & \qquad \left. \tilde{q}_2\left(m_2 \partial^2_t+\kappa_2\lambda \right) q_2 -\frac{D_2}{2}\tilde{q}_2^2 -\lambda\tilde{q}_2q_1 \right] \ .
\end{split}
\end{equation}

It is always possible to extend the upper limit of integration to $+ \infty$ since observables in unconditional stochastic processes are independent of future evolution. 
At the same time, if we are interested only in the properties of the steady-state, we can send $t_0 \to-\infty$. The path integral prepares the steady-state in such a limit starting irrespective of the initial state, meaning that the latter can be dropped without loss of generality (we can for example imagine the system starts diffusing from a delta-function on $q_1=q_2=0$ and zero momenta). In systems where the steady-state does not exist, one needs to be careful about the initial state (see~\cite{oppenheim2025diffusion} for an example).

To compute the unequal time 2-point function of the positions of the two oscillators, being the path integral itself Gaussian, it suffices to invert the kinetic matrix in the MSR action. Indeed:
\begin{equation}
    S[z] = \frac{1}{2}\int dt z^i A_i^j z_j \ ,
\end{equation}
and, for a Gaussian process, one has:
\begin{equation}
    \mathbb{E}[z^i(t) z^j(t')] = [A^{-1}(t,t')]^i_j \equiv (G(t,t'))^i_j \ .
\end{equation}
It is easier to invert the operator in Fourier space and only then Fourier transform back into $t$-space. In frequency domain, the operator is given by:
\begin{equation}
\label{eq:g_inv_cl_w}
    \mathbf{G}^{-1}(\omega) = \begin{pmatrix}
0 & m_1 \omega^2 + i\alpha \omega -\kappa_1 -\lambda & 0 & \lambda \\
m_1 \omega^2 - i\alpha \omega -\kappa_1 - \lambda & -D_1 & \lambda  & 0 \\
0& \lambda & 0 & m_2 \omega^2 - \kappa_2-\lambda \\
\lambda & 0 & m_2 \omega^2 - \kappa_2-\lambda  & -D_2 
\end{pmatrix}  \ ,
\end{equation}
which can be easily inverted using standard formulas for block 2x2 matrices~\cite{lu2002inverses}. The non-zero components are:
\begin{align}
\label{eq:g_cl_w}
    G^1_1(\omega) &= \frac{D_1 (m_2 \omega^2-\kappa_2-\lambda)^2}{|D(\omega)|^2}  +\frac{\lambda^2 D_2}{|D(\omega)|^2} \\
    G^2_2(\omega) &= \frac{D_2 |-m_1 \omega^2+\kappa_1+\lambda+i\omega\alpha|^2}{|D(\omega)|^2}  +\frac{\lambda^2 D_1}{|D(\omega)|^2} \\
    G^1_2(\omega) &= \frac{\lambda D_1 (-m_2 \omega^2+\kappa_2+\lambda)}{|D(\omega)|^2} + \frac{\lambda D_2 (-m_1 \omega^2+i\omega \alpha+\kappa_1+\lambda)}{|D(\omega)|^2} =  {G_1^2}^* \\
    G^1_{\tilde{1}}(\omega) &= \frac{m_2 \omega^2-\kappa_2-\lambda}{D(\omega)^*} = {G_1^{\tilde{1}}}^* \\
    G^2_{\tilde{2}}(\omega) &= \frac{m_1 \omega^2-i\omega \alpha-\kappa_1-\lambda}{D(\omega)^*} = {G_2^{\tilde{2}}}^* \\
    G^2_{\tilde{1}}(\omega) &=G^1_{\tilde{2}}(\omega)= -\frac{\lambda}{D(\omega)^*} = {G_2^{\tilde{1}}}^* =  {G_1^{\tilde{2}}}^* \ , \\
\end{align}
where
\begin{equation}
\label{eq:den_w}
    D(\omega) = (m_1 \omega^2- i \alpha \omega -\kappa_1-\lambda)(m_2\omega^2-\kappa_2-\lambda)-\lambda^2 \ ,
\end{equation}
and
\begin{equation}
    D(\omega)^* = (m_1 \omega^2+ i \alpha \omega -\kappa_1-\lambda)(m_2\omega^2-\kappa_2-\lambda)-\lambda^2 \ ,
\end{equation}
that is, we conjugate the coefficient only, not the argument. Then  $|D(\omega)|^2\equiv D(\omega)D(\omega)^*$.

The inverse Fourier transform of these frequency-domain two-point function can be easily computed using contour integration, once the roots of the quartic equation with complex coefficients $D(\omega)=0$ are known. These are in principle possible to find analytically, but they are extremely complicated expressions. However, some general statements can be made without knowing the exact form of the solutions. First, for contour integration it is crucial to know the sign of their imaginary part. To see this, consider:
\begin{multline}
    D(\theta=i\omega) = 
    \theta^4 - \gamma_1 \theta^3 + \left(\omega_1^2 +\omega_2^2 + \frac{\lambda}{m_1}+\frac{\lambda}{m_2} \right) \theta^2- \gamma_1 \left( \omega_2^2+\frac{\lambda}{m_2}\right)  \theta+\omega_1^2 \omega_2^2 \\
    + \omega_2^2\frac{\lambda}{m_1}+ \omega_1^2\frac{\lambda}{m_2} = 0 \ ,
\end{multline}
which is the same quartic that appeared in Equation~\ref{eq:eig_eq}. We therefore already know that its roots $\theta_i$ have strictly positive real parts and that, consequently, the imaginary parts of $\omega$ are strictly negative. This means, on the other hand, that the imaginary parts of the solutions to $D(\omega)^*$ are strictly positive. As a consequence, no pole ever lies exactly on the real axis, making the use of Cauchy's residue theorem in the Fourier transform straightforward (no pole prescription is needed). 

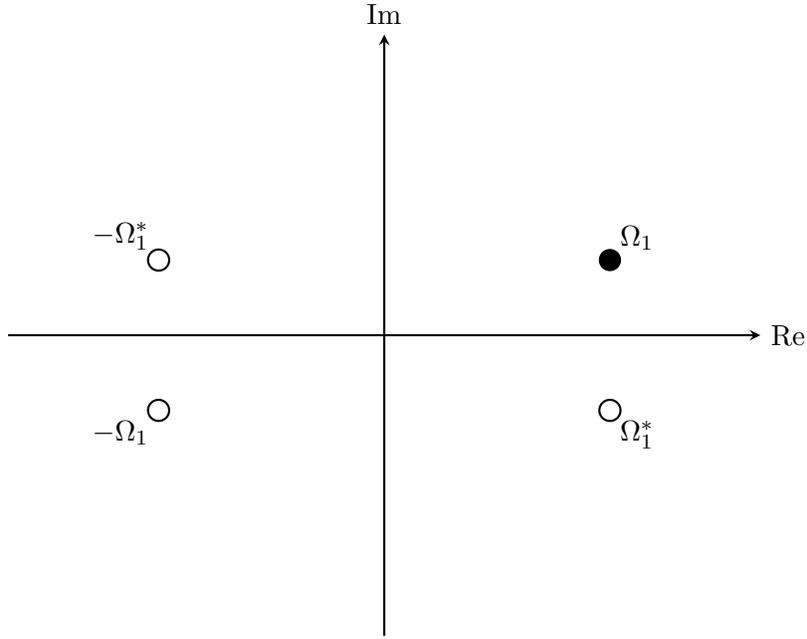
\begin{figure}[h]
    \centering
    \begin{tikzpicture}[>=stealth, scale=2]

        \draw[thick, ->] (-2.5,0) -- (2.5,0) node[right] {$\operatorname{Re}$};
        \draw[thick, ->] (0,-2) -- (0,2) node[above] {$\operatorname{Im}$};

        \def\q{1.5} 
        \def\y{0.5}   

        \filldraw[black] (\q,\y) circle (2pt) node[above right] {$\Omega_1$};

        \draw[black, thick] (-\q,-\y) circle (2pt);
        \node[below left] at (-\q,-\y) {$-\Omega_1$};

        \draw[black, thick] (-\q,\y) circle (2pt);
        \node[above left] at (-\q,\y) {$-\Omega_1^*$};

        \draw[black, thick] (\q,-\y) circle (2pt);
        \node[below right] at (\q,-\y) {$\Omega_1^*$};

    \end{tikzpicture}

    \caption{The $\Omega_1$ pole and its reflections in the complex plane.}
    \label{fig:pole_struct}
\end{figure}

The structure of the poles for the propagators is such that they can all be generated by two independent complex roots. Let's call:
\begin{equation}
    \Omega_1 = \tilde{\omega}_1 + i \tilde{\gamma}_1 \ \qquad \Omega_2 = \tilde{\omega}_2 + i \tilde{\gamma}_2
\end{equation}
the two independent roots of $D(\omega)^*$ living in the positive quadrant of the complex plane for some $\tilde{\omega}_1,\tilde{\omega}_2,\tilde{\gamma}_1,\tilde{\gamma}_2>0$. Then, we can generate all the other roots of both $D(\omega)$ and $D(\omega)^*$ and, consequently $|D(\omega)|^2$ by a combination of conjugation and reflection about the real axis:
\begin{itemize}
    \item $\Omega_1,\Omega_2,-\Omega_1^*,-\Omega_2^*$ are solutions to $D(\omega)^*$
    \item $\Omega_1^*,\Omega_2^*,-\Omega_1,-\Omega_2$ are solutions to $D(\omega)$ \ .
\end{itemize}
For convenience, we show the pictorial position in the complex plane of the poles in Figure~\ref{fig:pole_struct}.

It is useful to find the approximate roots for small coupling. In particular, we know the roots for $\lambda=0$ -- they are simply the eigenvalues of the two coupled systems:
\begin{equation}
   \Omega_{1,0} \equiv \Omega_1^{(\lambda=0)}=\sqrt{ \omega_1^2-\frac{\gamma_1^2}{4}} + i \frac{\gamma_1}{2} \ \qquad \Omega_{2,0} \equiv\Omega_2^{(\lambda=0)}=\omega_2 \ .
\end{equation}
Then, as we deform the system with $\lambda\neq0$, the roots will receive some small corrections, both real and imaginary (necessarily positively imaginary in the case of $\Omega_2$ as shown earlier). We can easily work out what that will be by expanding:
\begin{equation}
    \Omega_i = \Omega_i^{(0)} + \lambda \Omega_i^{(1)} + \lambda^2 \Omega^{(2)}_i + \mathcal{O}(\lambda^3) \equiv \Omega_i^{(0)}+\delta\Omega_i
\end{equation} \ 
and requiring $D(\Omega_i)=0$, to hold up to quadratic order in $\lambda$. We obtain:
\begin{align}
    \delta\tilde{\omega}_1 &=\frac{\lambda}{2m_1 \sqrt{\omega_1^2-\frac{\gamma_1^2}{4}}}\left[1+ \frac{\lambda}{ m_2\left((\omega_1^2-\omega_2^2)^2+\gamma_1^2 \omega_2^2 \right)}  \left(\omega_1^2-\omega_2^2-\frac{\gamma^2}{2}-\frac{m_2}{4m_1}\frac{\gamma^2 \omega_2^2+(\omega_1^2-\omega_2^2)^2}{\omega_1^2-\frac{\gamma^2}{4}} \right)\right] \\
    \delta\tilde{\gamma}_1 &= - \frac{\lambda^2}{2m_1m_2\left((\omega_1^2-\omega_2^2)^2+\gamma_1^2 \omega_2^2 \right) } \gamma_1
    \\
     \delta\tilde{\omega}_2 &=\frac{\lambda}{2m_2 \omega_2}\left[1 - \frac{\lambda}{ m_1\left((\omega_1^2-\omega_2^2)^2+\gamma_1^2 \omega_2^2 \right)} \left( \frac{m_1}{4m_2}\frac{\left((\omega_1^2-\omega_2^2)^2+\gamma_1^2 \omega_2^2 \right)}{\omega_2^2} +\omega_1^2- \omega_2^2\right)\right] \\
    \delta\tilde{\gamma}_2 &= \frac{\lambda^2}{2m_1m_2\left((\omega_1^2-\omega_2^2)^2+\gamma_1^2 \omega_2^2 \right) }\gamma_1  \ ,
\end{align}
A good sanity check is that the imaginary component of $\Omega_2$ is indeed positive. More importantly, corrections linear in $\lambda$ only shift the poles along the real axis, whilst the corrections to the imaginary components come only at second order in $\lambda$. Naturally, by small $\lambda$ we really mean that the frequencies associated with the interaction spring are much smaller than the natural frequencies of the two oscillators:
\begin{equation}
    \frac{\lambda^2}{m_1m_2} \ll \omega_1 \omega_2 \gamma_1^2 \ .
\end{equation}
For identical oscillators (i.e. $m_1 = m_2 = m_*$ and $\omega_1 = \omega_2 = \omega_*$) the corrections reduce to:
\begin{align}
    \delta\tilde{\omega}_1 &=\frac{\lambda}{2m_* \sqrt{\omega_*^2-\frac{\gamma_1^2}{4}}}\left(1- \frac{\lambda}{ 2 m_* \omega_*^2}  \left(1+ \frac{1}{2}\frac{\omega^2_*}{\omega_*^2-\frac{\gamma_1^2}{4}} \right)\right) \\
    \delta\tilde{\gamma}_1 &= - \frac{\lambda^2}{2m_*^2\gamma_1   \omega_*^2} 
    \\
     \delta\tilde{\omega}_2 &=\frac{\lambda}{2m_* \omega_*}\left(1 - \frac{\lambda}{4 m_* \omega_*^2 }  \right) \\
    \delta\tilde{\gamma}_2 &= \frac{\lambda^2}{2m_*^2\gamma_1   \omega_*^2} = -\delta\tilde{\gamma}_1  \ ,
\end{align}

To illustrate the behaviour of the system, we will focus on the small (yet finite) $\lambda$ limit. When performing the inverse Fourier transform of the two-point function, it is important to keep in mind that the residues of $|D(\omega)|^2$ for the $\Omega_2$ poles are of order $\lambda^{-2}$, as this is the scaling of the difference between such pole and their complex conjugate. This is since they have no finite imaginary part in the $\lambda\to 0$ limit. This will result in the two-point functions having some leading order $\lambda^{-n}$ terms -- a signature of the fact that for vanishing coupling the system does not have a steady-state. Again, it is in principle possible to compute the general solutions explicitly in terms of the roots of $D(\omega)$, but we refrain from reporting them as their are cumbersome and not particularly illuminating.

Performing the Fourier transform and keeping only terms leading order in $\lambda$ that do not vanish in the $\lambda\to0$ limit we obtain the following non-vanishing correlators:
\begin{align}
\label{eq:g_cl_t}
    G^1_1(t) &= \frac{1}{2\gamma_1 \omega^2_* m_*^2} \left[D_1 e^{-\frac{\gamma_1}{2}|t|} \left(\cos\left( \sqrt{\omega_*^2-\frac{\gamma_1^2}{4}} |t|\right) \right. \right. \\
    &\hspace{50pt} \left. \left. + \frac{\gamma_1}{2 \sqrt{\omega_*^2-\frac{\gamma_1^2}{4}}}\sin\left( \sqrt{\omega_*^2-\frac{\gamma_1^2}{4}} |t|\right) \right) 
     + 
D_2\cos(\omega_* |t|)\right] \notag \\
    G^2_2(t) &= \frac{D_2}{2} \frac{\gamma_1}{\lambda^2} \cos(\omega_* |t|)\\
    G^1_2(t) &= \frac{D_2}{2} \frac{1}{\lambda \omega_* m_*} \sin(\omega_* |t|) = G^2_1(t)\\
    G^1_{\tilde{1}}(t) &= \frac{1}{m_*} \frac{e^{\frac{\gamma_1}{2}t}}{\sqrt{\omega_*^2-\frac{\gamma_1^2}{4}}}\sin\left( \sqrt{\omega_*^2-\frac{\gamma_1^2}{4}} t \right) \theta(-t)= {G_1^{\tilde{1}}}(-t) \\
    G^2_{\tilde{2}}(t) &= \frac{1}{ m_* \omega_*} \sin(\omega_* t) \theta(-t)= {G_2^{\tilde{2}}}(-t) 
\end{align}
where we have defined
\begin{equation}
    G^i_j(t) = \mathbb{E}[z^i(\tau)z^j(\tau+t)] \ .
\end{equation}
An important check is that they match the equal-time covariances match Equations~\ref{eq:lyapunov_solp1p1} to~\ref{eq:lyapunov_solx1x2}, which were obtained from solving the Lyapunov equation instead. Note that the covariance between $q_1$ and $q_2$ in Equation~\ref{eq:lyapunov_solx1x2} goes to zero in the case of equal mass and coupling. The reader can find the general solution at small $\lambda$ in Appendix~\ref{app:gen_sol},

A few interesting properties emerge. First of all, the ratio between the typical size of the oscillations in the first spring with respect to the second one depends linearly on the coupling constant:
\begin{equation}
    \frac{\sigma_1}{\sigma_2} = \sqrt{\frac{G_1^1(0)}{G_2^2(0)}} = \frac{\lambda}{\gamma_1 m_*} \sqrt{1+\frac{D_1}{D_2}}
\end{equation}
Secondly, and more curiously, the covariance between the displacement of the first oscillator at different times for time intervals greater than the typical decay time of its fluctuations $t \gtrapprox 1/\gamma_1$ is long-lived and completely dominated by effects due to the second oscillator even at leading order in $\lambda$:
\begin{equation}
    G^1_1(t) \to \frac{D_2}{2}\frac{1}{\gamma_1 \omega_*^2 m_*^2} \cos(\omega_* |t|) \ .
\end{equation}
This reflects a key behaviour of the steady-state: the two oscillators synchronise, oscillating at the same frequency (albeit with in general a phase-difference). This is even more obvious when one looks at the mutual information between $q_1$ and $q_2$. Recalling that the probability distribution on the combined state is Gaussian, this is trivially given by:
\begin{equation}
    I_{ij}(t) \equiv I(x_i(\tau),x_j(\tau+t)) = -\frac{1}{2}\log\left(1-r_{ij}(t)^2 \right) \ ,
\end{equation}
where $r_{ij}$ is the correlation  between the two sytems:
\begin{equation}
    r_{ij}(t) = \frac{C^i_j(t)}{\sqrt{C^i_i(0) C^j_j(0)}} \ . 
\end{equation}
where the correlation functions are explicitly given by:
\begin{align}
    r_{11} &= \frac{1}{1+\frac{D_1}{D_2}}\left[ \cos(\omega_*t) + \frac{D_1}{D_2} e^{-\frac{\gamma_1}{2}|t|} \left(\cos\left( \sqrt{\omega_*^2-\frac{\gamma_1^2}{4}} |t|\right) \right. \right. \\ & \hspace{50pt} +\left. \left. \frac{\gamma_1}{2 \sqrt{\omega_*^2-\frac{\gamma_1^2}{4}}}\sin\left( \sqrt{\omega_*^2-\frac{\gamma_1^2}{4}} |t|\right) \right)\right] \notag \\
    r_{22} &= \cos(\omega_* t) \\
    r_{12} & = \frac{1}{\sqrt{1+ \frac{D_1}{D_2}}} \cos(\omega_* t) \ .
\end{align}

It is interesting to focus on the behaviour of the mutual information for the displacement of the damped oscillator at different times. Initially, it decays in magnitude until it asymptotes an oscillatory behaviour:
\begin{equation}
    \lim_{t\to \infty}I_{11} = -\frac{1}{2}\log\left[1-\frac{\cos^2(\omega_* t)}{\left(1+\frac{D_1}{D_2}\right)^2} \right] \ .
\end{equation}
This shows that information is scrambled about the position of the first oscillator after the half-life $1/\gamma_1$. After that, however, the mutual information between two observations oscillate between $0$ and some positive value set by the ratio of the two diffusion coefficients with period $\omega_*$. A similar functional dependence appears in $I_{12}$. This elucidates the fact that whilst the damped oscillator synchronises and vibrates at the natural frequency of the frictionless one in the steady-state, its dynamics is less regular that its counterpart. That is, if the diffusion coefficient of the first oscillator is large enough. Indeed, in the $D_1/D_2 \to 0$ limit, both $I_{11}$ and $I_{12}$ asymptote to $I_{22}$ -- the relative states of the two oscillators in the steady-state become completely deterministic.

\begin{figure}[h]
\centering
\includegraphics[width=\textwidth]{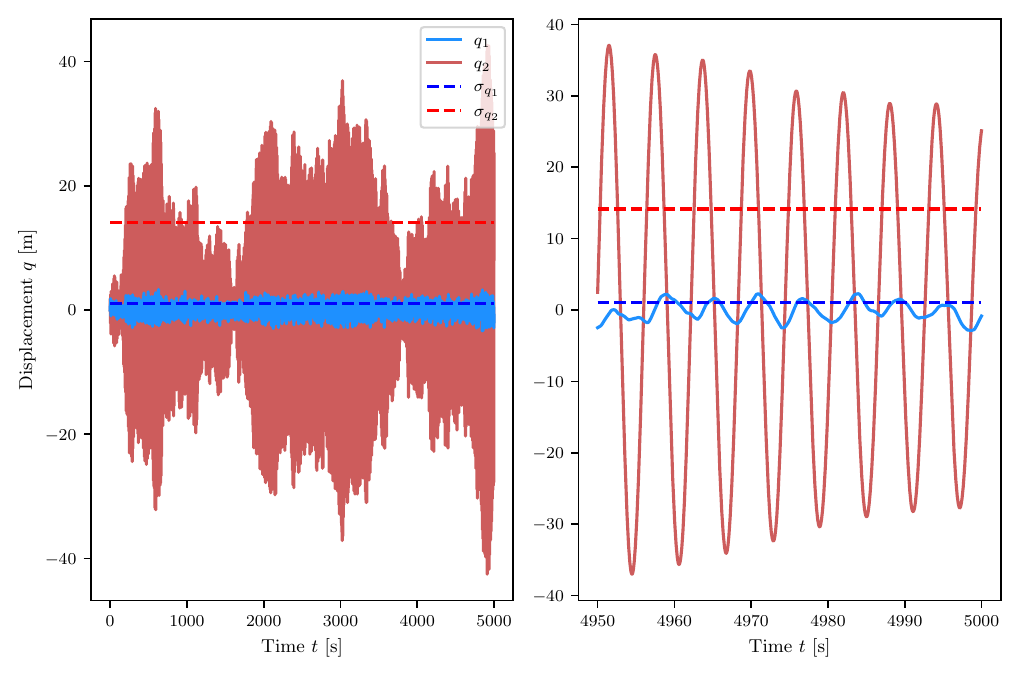} 
\caption{Sample trajectory for the couple oscillators with $m_*=1 \textrm{Kg}$, $\omega_*=\gamma_1=1 \textrm{s}^{-1}$ and $\lambda=0.05 \text{Nm}^{-1}$. Dashed lines are the theoretical predictions for the variance in the positions of the two oscillators. The numerical integration was performed with an Euler-Maruyama forward scheme with timestep $\Delta t=0.0005 \textrm{s}$.} \label{fig:traj_1}

\end{figure}

A sample trajectory for the coupled stochastic oscillators is shown in Figure~\ref{fig:traj_1}, where the theoretical predictions for the variance in the positions of the oscillators are shown. Indeed, one qualitatively sees that the system quickly reaches a steady state with the properties outlined above -- with the two masses oscillating with the same frequency, albeit with a relative phase.

\subsection{The classical-quantum case}
\label{sec:quantum}
Having analysed the system when both oscillators are classical, from the existence of the steady-state to its properties in the small coupling regime, we now study the classical-quantum case. In particular, we quantise the frictionless oscillator, in the attempt to answer the question of whether classical friction is enough for the combined CQ system to reach a steady-state. This is of interest especially in the case of effective CQ theories, where the quantum system is well isolated except for the interaction with the classical one, whose classicality is effective and comes from the interaction with some bath. If the latter is thermal, fluctuation-dissipation relations imply the existence of classical friction.

We tackle the problem from the path-integral formulation of the dynamics. For a classical oscillator with displacement $q$ coupled to a quantum one with displacement $Q$, we have that the proto-action encoding the CQ interaction is given by:
\begin{equation}
    \mathcal{W}_{CQ} = - \frac{\lambda}{2}(q-Q)^2 \ .
\end{equation}
Then, in the $L/R$ basis for the quantum system, the action $I_{CQ}$ for the hybrid path integral is given by:
\begin{multline}
    I_{CQ} = \int_0^T dt \left[i\left(\frac{1}{2}m_Q \left(\dot{Q}_L^2-\dot{Q}_R^2\right) - \frac{1}{2}\kappa_Q \left(Q_L^2 -Q_R^2 \right) -\left(\frac{\lambda}{2}(q-Q_L)^2
     - \frac{\lambda}{2} (q-Q_R)^2\right)\right)  \right. \\
     \left. - \frac{D_0}{2}\lambda^2(Q_L-Q_R)^2  
    -\tilde{q}(m_C \partial_t^2+\alpha \partial_t+\kappa_C +\lambda) q+ \frac{D}{2}\tilde{q}^2 + \frac{\lambda}{2}\tilde{q}(Q_L+Q_R)  \right] \ ,
\end{multline}
with $m_{Q,C}$ and $\kappa_{Q,C}$ being respectively the masses and coupling constants of the classical and quantum springs. As before, $\lambda$ is the coupling constant between the two oscillators and $\alpha$ the friction coefficient of the classical system. Finally, $D$ and $D_0$ are, respectively, the diffusion coefficient for the classical oscillator and the decoherence rate in the quantum one.

It is useful to expand the coupling term in the unitary part of the quantum action:
\begin{multline}
    I_{CQ} = \int_0^T dt \left[i\left(\frac{1}{2}m_Q \left(\dot{Q}_L^2-\dot{Q}_R^2\right) - \frac{1}{2}\kappa_Q \left(Q_L^2 -Q_R^2 \right) + \lambda q (Q_L-Q_R)\right) - \frac{D_0}{2}\lambda^2(Q_L-Q_R)^2   \right. \\
     \left.  
    -\tilde{q}(m_C \partial_t^2+\alpha \partial_t+\kappa_C +\lambda) q+ \frac{D}{2}\tilde{q}^2 + \frac{\lambda}{2}\tilde{q}(Q_L+Q_R)  \right] \ .
\end{multline}

It is suggestive that only the average of the left and right branches of the path integral acts as a source to the classical system, whilst the difference appears to couple to the classical system in the quantum sector of the path integral. Indeed, moving to the average-difference basis (suggestively also known as the classical-quantum basis, but we'll avoid that nomenclature to minimise confusion):
\begin{equation}
    Q_+ = \frac{Q_L + Q_R}{2} \ , \qquad Q_- = Q_L- Q_R \ ,
\end{equation}
one obtains the suggestive action:
\begin{multline}
\label{eq:ICQ}
    I_{CQ} = \int_0^T dt \left[-i Q_-\left(m_Q\partial_t^2 + \kappa_Q + \lambda \right)Q_+ - \frac{D_0}{2}\lambda^2q_-^2 + i \lambda q Q_-  \right. \\
     \left. 
    -\tilde{q}(m_C \partial_t^2+\alpha \partial_t+\kappa_C +\lambda) q+ \frac{D}{2}\tilde{q}^2 + \lambda\tilde{q}Q_+  \right] \ .
\end{multline}
where we have integrated by parts the kinetic term in the unitary sector of the action. This transformation elucidates a symmetry between the classical and quantum sectors of the hybrid system. Indeed, recall that the response variable $\tilde{q}$ in the MSR formalism is a purely imaginary auxiliary field. Making it explicit via the transformation $\tilde{q}\to i \tilde{q}$, we see that the average degree of freedom $Q_+$ is in exact correspondence with the classical displacement $q$, whilst $Q_-$ plays the role of the response variable. Of course, this is just a mathematical equivalence in the propagator of the theory: the reduced states of the classical and quantum systems will be a probability distribution and a density matrix respectively.

This arises due to a well-known equivalence between Lindblad evolution and Fokker-Planck equations in the case of Gaussian-preserving dynamics. Indeed, for quadratic potentials and Lindblad operators at most linear in $P$ and $Q$ (where $P$ is the conjugate momentum of a quantum particle with position $Q$) the evolution of the Wigner quasi-probability distribution representing the state of the system in phase-space follows exactly a Fokker-Planck-like equation\cite{isar1996phase,arnold2008quantum}. 

Introducing anharmonicities breaks this nice symmetry bewteen diffusive and Lindbladian dynamics. This is since, as se show explicitly in Section~\ref{sec:phase_space}, the quantum sector of the dynamics can be mapped exactly to a classical stochastic processes (modulo constraints on the initial state) if and only if the potential is at most quadratic in the generalised position of the system. As soon as the potential has a power expansion that goes beyond the quadratic term, such a mapping becomes at best approximative -- and only allowed in a region of phase space where the potential is effectively harmonic. Still, when the path integral is Gaussian, the Hubbard-Stratonovich transformation allows for an exact mapping between the classical stochastic evolution and the Lindbladian one. This is a powerful result, as it allows to use the properties of the equivalent diffusive generator to compute the steady-state of the quantum system~\cite{maghrebi2016nonequilibrium}. We will return to this point more formally in Section~\ref{sec:phase_space}.

This simplifies greatly the problem: we can use all the results from our classical-classical system, under the mapping
\begin{equation}
\tilde{q}_2 \to i Q_- \ , \qquad q_2 \to Q_+   \ , \qquad  D_2 \to D_0 \lambda^2 \ ,\qquad  D_1 \to D .
\end{equation}
If the backreaction is non-zero ($\lambda\neq 0$), the decoherence diffusion trade-off requires $4DD_0\geq 1$. We choose to saturate the trade-off setting $D_0=1/4D$ -- the special case of hybrid dynamics in which the quantum state remains pure conditioned on the classical trajectory~\cite{layton2024a-healthier}.
Most importantly, we can conclude that the combined system reaches a steady-state, meaning that we can extend the limits of integration in the CQ action of Equation~\ref{eq:ICQ} to past and future infinity, preparing the asymptotic state. What changes is the interpretation of the correlators and how they map to physical observables.

\subsubsection{Occupation number}
Much like in the classical systems, the correlations between $Q_-$ and $Q_+$ encode both correlations and the response of the system to external perturbations. In particular~\cite{sieberer2016keldysh}:
\begin{align}
    \langle Q_+(t)Q_+(t')\rangle & \equiv G^K(t,t') \ , \\
    \langle Q_-(t)Q_+(t')\rangle & \equiv iG^A(t,t') \ , \\
    \langle Q_+(t)Q_-(t')\rangle & \equiv iG^R(t,t') \ , \\
    \langle Q_-(t)Q_-(t')\rangle &= 0 \ .
\end{align}
The fact that the insertion of the difference field $Q_-$ computes the perturbation to the system due to the external source can be understood in terms of the observation that a real external source is physical and therefore equal on the $L$ and $R$ branches. However, by performing the Keldysh rotation at the level of the source, it is straightforward to see that $Q_+$ couples to the difference of the sources $J_-$ and vice-versa. Therefore, differentiating with respect to the physical source brings down a factor of $Q_-$.

Whilst the off-diagonal components of the quantum Green's function  encode the response of the system to external perturbations, the Keldysh propagator $G^K$ encodes the correlations in the system. In particular, introducing the usual bosonic raising $a^\dagger$ and lowering $a$ operators, it is straighforward to see that the equal-time Keldysh Green's function computes the average occupation of the oscillator:
\begin{equation}
    G^K(t,t) = \frac{1}{m_Q\omega_Q}\left(N+\frac{1}{2}\right) \ ,
\end{equation}
with $N$ being the expectation value of the number operator $a a^\dagger$.

The non-zero correlator $\langle\langle\tilde{q}Q_+ \rangle \rangle$ (where double-angled brackets indicate quantum \emph{and} classical expectation value) econde instead the response in the quantum degrees after a perturbation to the classical system, and vice-versa for $\langle\langle q Q_- \rangle \rangle$. The decoherence-diffusion trade-off forced the decoherence coefficient to have quadratic dependence on $\lambda$, meaning that with respect to the classical results the relative weights of certain terms is shifted towards the ones involving the classical oscillator only. Indeed, keeping only leading terms up to order $\lambda^0$, the non-zero correlators are given by:
\begin{align}
\label{eq:CQ_corr}
    \langle\langle q(0)q(t)\rangle\rangle &= \frac{1}{2\gamma_1 \omega^2_* m_*^2} \left[D e^{-\frac{\gamma_1}{2}|t|} \left(\cos\left( \sqrt{\omega_*^2-\frac{\gamma_1^2}{4}} |t|\right) \right.\right. \\
    &  \hspace{50pt}+ \left. \left. \frac{\gamma_1}{2 \sqrt{\omega_*^2-\frac{\gamma_1^2}{4}}}\sin\left( \sqrt{\omega_*^2-\frac{\gamma_1^2}{4}} |t|\right) \right)\right] \notag  \\
    \langle\langle Q_+(0)Q_+(t)\rangle\rangle &= \left(\frac{\gamma_1 }{8D}+\frac{D}{2\gamma_1 \omega_*^2 m_*^2} \right)\cos(\omega_* |t|)\\
    \langle\langle q(0)\tilde{q}(t)\rangle\rangle&= \frac{1}{m_*} \frac{e^{\frac{\gamma_1}{2}t}}{\sqrt{\omega_*^2-\frac{\gamma_1^2}{4}}}\sin\left( \sqrt{\omega_*^2-\frac{\gamma_1^2}{4}} t \right) \theta(-t) \\
    \langle\langle Q_+(0)Q_-(t)\rangle\rangle&= - \frac{i}{m_*} \sin(\omega_* t) \theta(-t) 
\end{align}
Note that, contrary to the classical-classical case, there are no divergences when $\lambda \to 0$, since saturating the decoherence-diffusion trade-off implies that the decoherence (and hence the energy increase) in the quantum state vanishes when the two systems decouple.

The energy in the quantum system is independent (to leading order) of the coupling constant between the two oscillators -- again a result of the decoherence-diffusion trade-off. Specifically, in analogy with the classical-classical case, the typical size of the oscillation in the quantum system would be $Q_+^2 \sim D_0/\lambda^2$ (since the induced friction is of quadratic order in the coupling constant), where $D_0$ is the decoherence strength -- effectively the diffusion constant in the quantum oscillator. However, the decoherence diffusion trade-off forces $D_0 \sim \lambda^2$, meaning that the two dependences on $\lambda$ cancel each other, giving an order 1 effect irrespective of the coupling strength. We see that two terms contribute to its average energy (essentially $Q_+^2$, the Keldysh propagator at equal times). The first one is direct decoherence in the system, controlled by $1/D$; the second one is linear in $D$ and is a result of the \emph{secondary} decoherence coming from the diffusion in the classical oscillator. That the combined decoherence and diffusion effects  produce an order 1 effect that cannot be hidden away is a powerful feature of CQ dynamics, making it an experimentally testable theory when applied to fundamental high-energy physics~\cite{oppenheim2023gravitationally,carney2024classical-quantum} Defining the effective temperature of the classical system to be:
\begin{equation}
    T_C=\frac{D}{2\alpha}=\frac{D}{2\gamma_1m_*}
\end{equation}
we see that we can re-express the average number of excitations in the quantum system as
\begin{equation}
   N = \frac{1}{2}\left( \frac{\omega_*}{2T_C}+ \frac{2T_C}{\omega_*}-1\right) \ .
\end{equation}
The first thing to note is that the quantum oscillator can never be empty of excitations. Tuning the temperature of the classical system to be the critical value $T_C^{crit}=\omega_*/2$ we can drive the quantum system to the lowest energy configuration allowed, namely the one that has $N_{min}=1/2$.
In the large diffusion regime (i.e. when the classical oscillator is much hotter than the zero-point energy of the quantum one), the direct decoherence is negligible and the energy in the classical and quantum oscillators exactly match to leading order in $\lambda$. In fact, we have that $N\approx T_C/\omega_*$, meaning that the quantum oscillator thermalises to $T_C$ as well.

We have used the correlations computed from the MSR path integral to find the two-point functions of the hybrid system to leading order in $\lambda$ (again, the system is in principle exactly solvable, but the roots of the quartic are extremely complicated and not at all illuminating). However, if we are only interested in equal-time \emph{correlations} -- that is we only care about symmetrised observables in the quantum system at equal times -- we can use the exact covariance computed from Equation~\ref{eq:lyapunov} (after appropriate rescalings of the coefficients). To see that the evolution of the average observables $Q_+$ is described by Equation~\ref{eq:class_eom}, we need to find the equation of motion for the conjugate momentum $P_+$. It suffices to take the momentum part of the full Schwinger-Keldysh action (the purely quantum sector of the CQ action) and perform the rotation in the average-difference basis before integrating out $P$:
\begin{equation}
\begin{split}
    S_{SK}[Q_L,P_L,Q_R,P_R] &= i \left[P_L \dot{Q}_L - \frac{P_L^2}{2m_Q}-P_R \dot{Q}_R + \frac{P_R^2}{2m_Q} \right] +... \\
    &= i\left[P_- \left(\dot{Q}_+ - \frac{P_+}{m_Q} \right)+P_+ \dot{Q}_- \right]+... \ .
\end{split}
\end{equation}
Integration over $P_-$ then sets $P_+=m_Q \dot{Q}_+$, and complete equivalence follows.

The non-zero equal-time two-point functions of the hybrid state are given by:
\begin{align}
    \langle \langle p^2 \rangle \rangle &= \frac{1}{2\gamma_1}\left(D + \frac{m_C}{m_Q} \frac{\lambda^2}{4D} \right) \\
    \langle \langle P^2 \rangle \rangle &=  \frac{1}{2\gamma_1} \left[\frac{\lambda^2}{4D}\left(1 + \frac{m_C m_Q}{\lambda^2} \left(\left(\omega_C^2-\omega_Q^2 + \frac{\lambda}{m_C} - \frac{\lambda}{m_Q} \right)^2 \right. \right. \right. \\
    &  \hspace{50pt} \left. \left. \left. + \gamma_1^2 \left(\omega_Q^2 +\frac{\lambda}{m_Q} \right)\right) \right) 
      +\frac{m_Q}{m_C} D \right] \notag  \\
    \langle \langle q^2 \rangle \rangle &= \frac{1}{2\gamma_1} \frac{D\left(\frac{\lambda}{m_Q}+\omega_Q^2\right)+\frac{m_C}{m_Q}\frac{\lambda^2}{4D}\left(\frac{\lambda}{m_C}+\omega_C^2\right)}{m_C^2 \left(\omega_Q^2 \frac{\lambda}{m_C}+\omega_C^2 \left(\omega_Q^2 + \frac{\lambda}{m_Q} \right) \right)} \\
    \langle \langle Q^2 \rangle \rangle &= \frac{1}{2\gamma_1} \frac{1}{m_Q^2 \left(\omega_Q^2 \frac{\lambda}{m_C}+\omega_C^2 \left(\omega_Q^2 + \frac{\lambda}{m_Q} \right) \right)} \left[\frac{m_Q}{m_C}D \left(\omega_C^2+\frac{\lambda}{m_C}\right) \right. \\
    & \hspace{60pt} +\frac{m_Cm_Q}{4D}\left(\left(\omega_C^2 +\frac{\lambda}{m_C}\right)^3+ \left(\omega_C^2 \left( \omega_Q^2+\frac{\lambda}{m_Q}\right) +\omega_Q^2 \frac{\lambda}{m_C}\right) \right. \notag \\
    & \left. \left.  \hspace{60pt} \times \left(\omega_Q^2 -2\omega_C^2+\frac{\lambda}{m_Q}-2\frac{\lambda}{m_C}+\gamma_1^2\right)\right) \right] \notag  \\ 
    \langle \langle q Q \rangle \rangle &= \frac{1}{2\gamma_1}\frac{m_C}{m_Q} \frac{D \frac{\lambda}{m_C}+\frac{m_C\lambda}{4D}\left(\left(\omega_C^2+\frac{\lambda}{m_C}\right)^2 -\omega_Q^2 \frac{\lambda}{m_C}-\omega_C^2 \left(\omega_Q^2+\frac{\lambda}{m_Q} \right)\right) }{\omega_Q^2\frac{\lambda}{m_C}+\omega_C^2\left(\omega_Q^2+\frac{\lambda}{m_Q} \right)} \\
    \langle \langle P q \rangle \rangle &= -\frac{\lambda}{8D} \\
    \langle \langle p Q \rangle \rangle &= \frac{\lambda}{8D} \frac{m_C}{m_Q} \\
    \langle \langle pP \rangle \rangle &= \frac{\lambda}{8 D\gamma_1} m_C \left( \omega_C^2-\omega_Q^2+\frac{\lambda}{m_C}-\frac{\lambda}{m_Q}\right) \ .
\end{align}

\subsubsection{Thermal limit}
In~\cite{layton2025restoring}, the temperature-dependent hybrid dynamics that preserves the CQ thermal state at any $\beta$ was derived -- and a CQ oscillator was studied as a toy model. In that work, the authors find that, in order to preserve the thermal state, a temperature-dependent decoherence in $P$ is required. Still, in the high-temperature limit the momentum decoherence term drops out, and their dynamics coincides with ours -- meaning that the model we discuss must flow to the CQ thermal state in the high-temperature regime as well -- as we now straightforwardly show.

A large effective temperature for the classical system at fixed $\gamma_1$ corresponds to the high diffusion limit. At large $D$, the non-zero 2-point functions are
\begin{align}
    \langle \langle p^2 \rangle \rangle &= m_C T_C  \\
    \langle \langle P^2 \rangle \rangle &=  m_Q T_C \\
    \langle \langle q^2 \rangle \rangle &= \frac{T_C}{m_C} \left(\frac{\omega_Q^2 \frac{\lambda}{m_C}}{\frac{\lambda}{m_Q}+\omega_Q^2}+\omega_C^2 \right)^{-1} \\
    \langle \langle Q^2 \rangle \rangle &= \frac{T_C}{m_Q} \left(\frac{\omega_C^2 \frac{\lambda}{m_Q}}{\frac{\lambda}{m_C}+\omega_C^2}+\omega_Q^2 \right)^{-1} \\
    \langle \langle q Q \rangle \rangle &= \frac{T_C}{m_Q} \left(\omega_Q^2+\frac{m_C\omega_C^2}{\lambda}\left(\omega_Q^2+\frac{\lambda}{m_Q} \right)\right)^{-1}
\end{align}

It is straightforward to see that in the high temperature regime the correlations converge exactly towards those of the thermal state:
\begin{equation}
    \varrho_\beta(q,p)=\frac{1}{\mathcal{Z}}e^{-\beta H(q,p)} \ ,
\end{equation}
with $\beta= 1/T_C$ as $\beta\to 0$. In that limit the hybrid thermal state limits the classical one, and the correlations can be easily extracted from the Gaussian state without worrying about the discreteness of the energy levels in the quantum system. That is, the partition function of the quantum oscillator is well-approximated by the classical one.

\subsection{CQ in phase space}
\label{sec:phase_space}
The dynamical equivalence between the CQ and the CC stochastic oscillators is not a coincidence. As mentioned, it is just an extension of the statement that classical and quantum generators are equivalent for harmonic potentials. To see this more explicitly, let's introduce the phase-space description of CQ dynamics by peforming a Wigner-Moyal transform, in the spirit of~\cite{isar1996phase}. For simplicity, we restrict to minimal CQ dynamics of the form of Equation~\ref{eq:CQ_master}. We further take the CQ Hamiltonian to be Hermitian and the quantum degree of freedom being described by a single point-particle. Extensions to higher-dimensional Hilbert space and more general CQ evolution are conceptually trivial.

The Wigner-Moyal transform assigns to every classical phase-space dependent operator $\hat{A}(z)$ (we introduce hats for operators and powers of $\hbar$ in this section to minimise confusion) a function over the combined phase space $\mathcal{M}_C \times \mathcal{M}_Q$
\begin{equation}
    \mathcal{W}\left[\hat{A}(z) \right]=A(z,Q,P) = \int dZ e^{ipZ/\hbar} \langle Q-Z/2|\hat{A}(z)|Q+Z/2 \rangle \ ,
\end{equation}
where $P$ and $Q$ are the position and momentum respectively of the quantum particle, whilst $|Q\rangle$ is its position eigenstate with eigenvalue $Q$. The classical phase space dependence of the operators does not add any complication here. The Wigner-Moyal transform of the CQ state then corresponds to the hybrid Wigner quasi-probability distribution $W$
\begin{equation}
     \mathcal{W}\left[\frac{1}{2\pi\hbar}\hat{\varrho}(z) \right] = W(z,P,Q) \ ,
\end{equation}
where the numerical factor is needed to appropriately normalise the state, since
\begin{equation}
    \int dQ \int dP \ \mathcal{W}\left[\hat{A}(z) \right] = 2\pi \hbar \mathrm{Tr}[\hat{A}(z)]  \ .
\end{equation}
This is just the usual Wigner function. The twist is that it is subnormalised on the quantum phase-space, and it has classical-phase space dependence.

The time evolution of the hybrid phase-space state $W$ is given by the Wigner-Moyal transform of Equation~\ref{eq:CQ_master}, the evolution map of the CQ state. In order to compute what that is in phase space, it is useful to keep in mind the following
\begin{equation}
\begin{split}
    \mathcal{W}\left[\hat{A}(z)\hat{B}(z) \right] &= A(z,Q,P) \exp\left(\frac{\hbar \Lambda}{2 i}\right)B(z,Q,P)\\ 
    &= B(z,Q,P) \exp\left(-\frac{\hbar \Lambda}{2 i}\right)A(z,Q,P) \ ,
\end{split}
\end{equation}
where the differential operator $\Lambda$ is essentially the negative of the Poisson brackets
\begin{equation}
    \Lambda = \partial_P^\leftarrow\partial_Q^\rightarrow - \partial_Q^\leftarrow\partial_P^\rightarrow \ ,
\end{equation}
with the arrow indicating what the derivative acts on. It then follows that the Wigner transform of the commutator is (from now on we drop the phase space dependence for notational economy)
\begin{equation}
     \mathcal{W}\left[ [\hat{A}(z),\hat{B}(z)] \right] = -2i \  A \sin\left(\frac{\hbar \Lambda}{2 }\right)B \ ,
\end{equation}
whilst for the anticommutator we obtain
\begin{equation}
     \mathcal{W}\left[ \{\hat{A}(z),\hat{B}(z)\}_+ \right] = 2 \ A\cos\left(\frac{\hbar \Lambda}{2 }\right) B \ .
\end{equation}
Using these relations, we can easily see that the reversible part of the CQ evolution equation, the Aleksandrov brackets, gets mapped to
\begin{equation}
\label{eq:WM_Aleks}
   \frac{1}{2\pi\hbar} \mathcal{W}\left[\{H_{CQ} , \varrho\}_A \right] = \left\{H_C + V_I \cos\left(\frac{\hbar \Lambda}{2 }\right), W\right\} -\frac{2}{\hbar}H_Q\sin\left(\frac{\hbar \Lambda}{2 }\right) W
\end{equation}
Here we have often used that the Wigner-Moyal transformation commutes with derivatives with respect to the classical degrees of freedom $z$. Note that, in the $\hbar \to 0$ limit, this is exactly the classical Liouville equation
\begin{equation}
   \frac{1}{2\pi\hbar} \mathcal{W}\left[\{H_{CQ} , \varrho\}_A \right] = \left\{H_C + V_I +H_Q , W \right\} + \mathcal{O}(\hbar)^2 \ ,
\end{equation}
as required for consistency. For the minimal models we consider, that is where the CQ interaction potential only involves generalised positions of the hybrid system ($V_I=V_I(q,Q)$) and similarly the quantum Hamiltonian is given by $H_Q = P^2/2m_Q +V_Q(Q)$, we can expand Equation~\ref{eq:WM_Aleks} as
\begin{equation}
\begin{split}
    \frac{1}{2\pi\hbar} \mathcal{W}\left[\{H_{CQ} , \varrho\}_A \right] =& \left\{H_C + V_I +H_Q , W \right\} \\
    &+ \sum_{n=1}^\infty (-1)^n \left(\frac{\hbar}{2}\right)^{2n}\left[\frac{1}{2n!} \frac{\partial^{2n}}{\partial Q^{2n}}\left( \frac{\partial V_I}{\partial q^i}\right)\frac{\partial^{2n}}{\partial P^{2n}}\left( \frac{\partial W}{\partial p_i}\right) \right. \\
    &\left.+\frac{1}{(2n+1)!} \frac{\partial^{2n+1}U}{\partial Q^{2n+1}}\frac{\partial^{2n+1}W}{\partial P^{2n+1}}\right] \ .
\end{split}
\end{equation}
This explicitly shows that, if $H_{Q}+V_I$ is at most harmonic, the reversible part of the dynamics is equivalent to the classical evolution -- generalising the standard quantum result to CQ systems. This is since the tower of derivatives vanishes identically for any value of $\hbar$.

What about the dissipative contribution instead? The Wigner-Moyal representation of the diffusive term is trivial, again because the map commutes with the derivatives with respect to $z$. On the other hand, the decoherence term is essentially equivalent to what has been computed in~\cite{isar1996phase}, under the appropriate rescalings, modulo the classical phase-space dependence. Indeed it is easy to show that
\begin{equation}
\begin{split}
    \frac{1}{2\pi\hbar} \mathcal{W}\left[\mathbf{D}[ \varrho] \right] = \frac{1}{2}\frac{\partial^2}{\partial p_i \partial p_j}\left(D_{2,ij} \ W  \right) + 2 D_0^{ij} \frac{\partial V_I}{\partial q^i} \sin\left(\frac{\hbar \Lambda}{2 }\right) \frac{\partial V_I}{\partial q^j}\sin\left(\frac{\hbar \Lambda}{2 }\right) W \ .
\end{split}
\end{equation}
Again, this can be expanded in powers of $\hbar$ in terms of an infinite tower of derivatives (using the Cauchy product for the two infinite series coming from the sines) 
\begin{equation}
\begin{split}
    \frac{1}{2\pi\hbar} \mathcal{W}\left[\mathbf{D}[ \varrho] \right] =& \frac{1}{2}\frac{\partial^2}{\partial p_i \partial p_j}\left(D_{2,ij} \ W  \right) + \frac{\hbar^2 D_0^{ij} }{2} \frac{\partial^2V_I}{\partial Q\partial q^i}\frac{\partial^2V_I}{\partial Q\partial q^j} \frac{\partial^2 W}{\partial P^2}\\
    & + 2D_0^{ij} \sum_{n=1}^\infty \sum_{m=0}^n \frac{(-1)^n}{c_{m,n}}\left( \frac{\hbar}{2}\right)^{2n+2} \frac{\partial^{2m+2}V_I}{\partial Q^{2m+1}\partial q^i} \frac{\partial^{2(n-m)+2}V_I}{\partial Q^{2(n-m)+1}\partial q^j} \frac{\partial^{2n+2}W}{\partial P^{2n+2}} \ ,
\end{split}
\end{equation}
where we defined $c_{m,n}\equiv(2m+1)!(2(n-m)+1)!$. We have explicitly isolated the $n=m=0$ component of the sum, since it obviously maps to a diffusion term under the Wigner-Moyal transform. Moreover, under the assumption of harmonic Hamiltonian, as before, that's the only term surviving. In which case we see again that the CQ master equation for quadratic potentials can be mapped exactly to a diffusion problem in phase space. This is indeed the parallel of what we have observed at the level of the path integral.

A word of caution: the decoherence-induced diffusion in the quantum system is negligible with respect to the classical one, unless the decoherence constant itself is of the order of $1/\hbar^2$. In effective open system, the induced decoherence rate is exactly of that order, meaning that in the $\hbar\to 0$, both effects contribute equally~\cite{zurek2003decoherence}. By inserting the explicit form of the potentials for the coupled CQ oscillators, and imposing the decoherence-diffusion trade-off, one can indeed see that in CQ the diffusion coefficient in the quantum variables is given by $\lambda^2\hbar^2/4D$, as in the discussion at the level of the path integral.

The phase-space description we have introduced here nicely mirrors the purely quantum-mechanical counterpart. Whilst it is an exact alternative representation of CQ dynamics, it can -- in analogy to the quantum case -- provide great computational advantage in evaluating the evolution of hybrid systems. For example, the Wigner formalism in quantum mechanics is useful when simulating molecular dynamics or highly transient phenomena~\cite{sellier2015an-introduction}.

\section{Discussion} 
\label{sec:discussion}
In this article, we explored a solvable system of classical-quantum interaction: the hybrid oscillator. We began with two classical stochastic oscillator, one of which experiencing friction, showing that such a system univocally flows to a non-equilibrium steady-state. We then computed the out-of-time correlators for the steady-state in the small coupling regime. Next, we quantised the undamped oscillators, and studied the system with the CQ framework. By mapping the generator of the dynamics to the classical stochastic system, we showed that also the hybrid oscillator flows to a non-equilibrium steady-state, which we computed. We demonstrated that in the high-diffusion regime of CQ, such a state becomes thermal. We concluded by formally deriving the phase-space description of CQ dynamics by performing the Wigner-Weyl transformation of the CQ generator. We showed explicitly that for quadratic potentials the hybrid evolution is equivalent to a Fokker-Planck equation with diffusion in both the classical and quantum phase space.

Classical friction can be enough for minimal hybrid system to reach a steady state, even though it will not be an equilibrium state in general. Whilst thermal states in CQ models have been recently studied in detail~\cite{layton2025restoring}, much can still be said on non-equilibrium states. In particular, the hybrid oscillator would be a good toy model to study properties of hybrid systems that do not satisfy detailed balance. A first objective would be to derive the generalised fluctuation relations for CQ non-equilibrium thermodynamics and compute the entropy production in the system \cite{landi2021irreversible,clarke2024stochastic}. 

Both classical and quantum thermal equilibrium and the fluctuation-dissipation relations can be shown to be associated with particular symmetries of the respective path-integral actions. A key step towards a complete understanding of CQ thermodynamics would be to show that such an equivalence exists for hybrid systems as well, deriving the fluctuation-dissipation relations from first principles in the process \cite{sieberer2015thermodynamic}. Another interesting avenue would be to make contact between quantum and hybrid thermodynamics, deriving the latter as a special case of the former -- possibly integrating out some environment à la Caldera-Leggett \cite{caldeira1981influence}.

\vspace{3mm} 

\noindent\section*{Acknowledgements}

The author would like to thank Jonathan Oppenheim, Isaac Layton, Muhammad Sajjad, Rhys Evans, Johannes Noller, Gautam Satishchandran, Harvey Weimberg and Lorenzo Braccini -- in random order -- for useful feedback and productive discussions.  
\newpage

\appendix

\section{Uncoupled oscillators}\label{app:uncoupled_osc}
\subsection{Non-dissipative oscillator}
Before moving onto the coupled system, it is useful to review the behaviour of a single stochastically driven damped and undamped oscillator. Let's start with the latter:
\begin{equation}
    m_2\ddot{q}_2 + \kappa_2 q_2  = \xi_2 \ ,
\end{equation}
i.e. an oscillator of natural frequency
\begin{equation}
    \omega_2 = \sqrt{\frac{\kappa_2}{m_2}}
\end{equation}
and no friction.
It is useful to express this second-order stochastic equation in terms of a first order system introducing the momentum $p_2$ of the particle:
\begin{equation}
\begin{split}
\label{eq:undamped}
    \dot{q}_2 - \frac{p_2}{m_2} & = 0 \\
    \dot{p}_2 + \kappa_2 q_2  & =\sqrt{D_2} \xi_2 \ .
\end{split}
\end{equation}
Note, this is an Ornstein-Uhlenbeck processes with
\begin{equation}
    \mathbf{\Theta} = \begin{pmatrix}
0 & -\frac{1}{m_2} \\
\kappa_2 & 0 
\end{pmatrix} 
\ , \qquad
    \mathbf{\Sigma} = \begin{pmatrix}
0 & 0 \\
0 & \sqrt{D_2} 
\end{pmatrix} \ .
\end{equation}

Intuitively, the system undergoes unbounded diffusion and heats up forever, which can be easily proven using It\^o's lemma. That is, the evolution of any function $f$ of a vector of stochastic variables obeying (Einstein's summation is assumed)
\begin{equation}
    \dot{z}^i = \mu^i + \sigma^i_j\xi^j \ , \qquad \mathbb{E}[\xi^i(t) \xi^j(t')] = \delta^{ij}\delta(t,t')
\end{equation}
is given by
\begin{equation}
    \dot{f} = \left(\mu^i \frac{\partial f}{\partial x^i}   + \frac{1}{2} \sigma^i_k \sigma^j_k \frac{\partial^2 f}{\partial x^i \partial x^j}\right) + \sigma^i_j \frac{\partial f}{\partial x^i} \xi^j .
\end{equation}

Applying it to the energy of the particle:
\begin{equation}
    H_2 = \frac{p_2^2}{2m_2}+\frac{1}{2}\kappa_2q_2^2  
\end{equation}
we obtain:
\begin{equation}
    \dot{H_2} = \frac{\sqrt{D_2}}{m_2}\xi_2 + \frac{D_2}{2m_2}  \ .
\end{equation}
Note for $D_2=0$ the energy of the system stays constant as expected, since the deterministic system is conservative. However, this means that the energy in the stochastic oscillator ($D_2 \neq 0$) is going to increase linearly in time on average:
\begin{equation}
    \mathbb{E}[H_2] = H_0+\frac{\sqrt{D_2}}{m_2}t \ .
\end{equation}

This is a signature that the system does not reach a steady-state, as easily checked using standard results from stochastic systems (and in agreement with expectations). For a OU process, a steady-state exists if and only if the deterministic system is strongly stable. For Equation~\ref{eq:undamped} this is clearly not the case, since the eigenvalues of $\mathbf{\Theta}$ are purely imaginary:
\begin{equation}
    \theta_{1,2}= \pm i\omega_2 \ .
\end{equation}

\subsection{Damped oscillator}
Adding any amount of damping to the system is enough for the oscillator to eventually reach a steady-state. Indeed, consider the damped stochastic oscillator in isolation (in first order representation):
\begin{equation}
\begin{split}
\label{eq:undamped}
    \dot{q}_1 - \frac{p_1}{m_1} & = 0 \\
    \dot{p}_1 + \frac{\alpha}{m_1} p_1 + \kappa_1 q_1  & =\sqrt{D_1} \xi_1 \ .
\end{split}
\end{equation}
Studying the evolution of the energy of the system suggests that indeed this system will have a steady-state:
\begin{equation}
    \dot{H_1}= - \frac{\alpha}{m_1^2}p_1^2 + \frac{\sqrt{D_1}}{m_1} \xi_1+ \frac{D_1}{2m_1} \ .
\end{equation}
Note that, here, we are only looking at the single damped oscillator.
Taking the expectation value of both sides, we can see that the average energy stops growing when:
\begin{equation}
    \text{Var}(p_1) = \frac{D_1}{2\gamma_1} \ .
\end{equation}
In principle, we can use this result together with the equations of motion to extract all the moments of the stochastic degrees of freedom. To show that the system does indeed reach a steady-state, however, it is quicker to note that this is still an OU process in the form of Equation~\ref{eq:OU} with
\begin{equation}
\label{eq:mat_damp}
    \mathbf{\Theta} = \begin{pmatrix}
0 & -\frac{1}{m_1} \\
\kappa_1 & \frac{\alpha}{m_1}
\end{pmatrix} 
\ , \qquad
    \mathbf{\Sigma} = \begin{pmatrix}
0 & 0 \\
0 & \sqrt{D_1} 
\end{pmatrix} \ .
\end{equation}
The eigenvalues of $\Theta$ are then:
\begin{equation}
    \theta_{1,2} = \frac{\gamma_1}{2} \pm \frac{1}{2} \sqrt{\gamma_1^2-4\omega_1^2} \ .
\end{equation}
They can be completely real or have an imaginary components (corresponding to the overdamped and underdamped regimes respectively), but they have positive real parts for any $\omega_1\neq0$ and $\gamma_1>0$, where
\begin{equation}
    \omega_{1} = \sqrt{\frac{\kappa_1}{m_1}} \ , \qquad \gamma_1 = \frac{\alpha}{m_1}  \ ,
\end{equation}
proving the existence of steady-state formally.

For an OU process, if the steady-state exists then it is Gaussian~\cite{godreche2018characterising}:
\begin{equation}
    P_{st}=(2\pi)^{-N/2} \mathrm{det}(\mathbf{C}_{\infty})^{-1/2} \exp\left(\frac{1}2{}z_i (C_{\infty}^{-1}) ^i_j z^j\right)
\end{equation}
The equal-time covariance of the OU process $\mathbf{C}_{\infty \ j}^i=\text{cov}(z^i,z^j)$ in such a state can be computed from the Lyapunov equation~\cite{godreche2018characterising}:
\begin{equation}
\label{eq:lyapunov_app}
    \mathbf{\Theta}\mathbf{C}_{\infty} + \mathbf{C}_{\infty} \mathbf{\Theta}^T = \mathbf{\Sigma} \mathbf{\Sigma}^T \ .
\end{equation}
For Equation~\ref{eq:mat_damp}, this can be readily solved giving:
\begin{equation}
\label{eq:damped_ss}
\mathbf{C}_\infty = \frac{D_1}{2 \gamma_1}\begin{pmatrix}
\frac{1}{m_1 \kappa_1} & 0 \\
0 & 1 
\end{pmatrix} \ ,
\end{equation}
which indeed matches the variance of the momentum we calculated earlier, and gives the spread in position for free as well. Equation~\ref{eq:damped_ss} is, of course, just the covariance associated with the thermal state:
\begin{equation}
    P(q_1,p_1) = \frac{1}{Z}e^{-\beta H_1} \ ,
\end{equation}
where $\beta=2\gamma_1m_1/D_1$, and $Z$ the appropriate normalisation factor.

\section{General solution}
\label{app:gen_sol}
In this appendix, we present the non-vanishing leading terms in the small coupling $\lambda$ expansion for the two-point functions of the positions of the classical stochastic oscillators. To compute them, it suffices to perform the Fourier transform of their frequency representation, making use of Cauchy residue theorem. Then, expanding the roots in powers of $\lambda$, one finds that the leading terms for the time-domain two-point functions are given by
\begin{align}
    G^1_1(t) &= \frac{D_1}{2\gamma_1 m_1^2 \omega_1^2} \left[e^{-\frac{\gamma_1}{2}|t|} \left(\cos\left( \sqrt{\omega_1^2-\frac{\gamma_1^2}{4}} |t|\right) \right. \right. \\
    &\hspace{50pt} \left. \left. + \frac{\gamma_1}{2 \sqrt{\omega_1^2-\frac{\gamma_1^2}{4}}}\sin\left( \sqrt{\omega_1^2-\frac{\gamma_1^2}{4}} |t|\right) \right) 
     + \frac{\omega_1^2}{\omega_2^2}\frac{m_1}{m_2}\frac{D_2}{D_1}
\cos(\omega_2 |t|)\right] \notag \\
    G^2_2(t) &= \frac{D_2}{2\gamma_1\omega_2^2} \frac{m_1}{m_2}\frac{(\omega_1^2-\omega_2^2)^2+\gamma_1^2 \omega_2^2}{\lambda^2} \cos(\omega_2 |t|)\\
    G^1_2(t) &= \frac{D_2}{2 m_2 \omega_2 \lambda} \left( \sin(\omega_2|t|)- \frac{\omega_2^2-\omega_1^2}{\gamma_1 \omega_2} \cos(\omega_2 |t|) \right) = G^2_1(t)\\
    G^1_{\tilde{1}}(t) &= \frac{1}{m_1} \frac{e^{\frac{\gamma_1}{2}t}}{\sqrt{\omega_1^2-\frac{\gamma_1^2}{4}}}\sin\left( \sqrt{\omega_1^2-\frac{\gamma_1^2}{4}} t \right) \theta(-t)= {G_1^{\tilde{1}}}(-t) \\
    G^2_{\tilde{2}}(t) &= \frac{1}{ m_2 \omega_2} \sin(\omega_2 t) \theta(-t)= {G_2^{\tilde{2}}}(-t) \\
    G^2_{\tilde{1}}(t) &= \mathcal{O}(\lambda)
\end{align}
These two point functions reduce to the quoted expressions in the main body when taking the limits $m_1 \to m_2 \equiv m_*$ and $\omega_1 \to \omega_2 \equiv \omega_*$.

\bibliography{comprehensive}

\end{document}